\documentclass[aps,prd,preprintnumbers,superscriptaddress,showpacs,nofootinbib,11pt,floatfix]{revtex4}
\usepackage{graphicx}
\usepackage{dcolumn}
\usepackage{bm}
\usepackage{amsmath}
\usepackage{amssymb}

\def\beq{\begin{equation}}
\def\eeq{\end{equation}}
\def\[{\left[}
\def\]{\right]}

\usepackage[large]{subfigure}
\usepackage{wrapfig}

\newcommand{\ole}{\overline}
\newcommand{\nn}{\nonumber}
\newcommand{\newc}{\newcommand}
\newc{\gsim}{\lower.7ex\hbox{$\;\stackrel{\textstyle>}{\sim}\;$}}
\newc{\lsim}{\lower.7ex\hbox{$\;\stackrel{\textstyle<}{\sim}\;$}}
\begin{document}
\title{Flavor-changing neutral current in production
and decay of pseudoscalar mesons in a type III
two-Higgs-doublet-model with four-texture Yukawa couplings}
\author{M. G\'omez-Bock$^1$, G.~ L\'opez-Castro$^2$, L.~
L\'opez-Lozano$^3$, A.~ Rosado$^3$ \\
\small $^1$Instituto de F\'{\i}sica, Universidad Nacional Aut\'onoma
de M\'exico,\\
\small Apartado Postal 20-364, M\'exico 01000 D.F., M\'exico\\
\small $^2$Departamento de F\'{\i}sica, Cinvestav,\\
\small Apartado Postal 14-740, M\'exico 07000 D.F., M\'exico.\\
\small $^3$Instituto de F\'{\i}sica, Benem\'erita Universidad
Aut\'onoma de Puebla. \\
\small Apartado Postal J-48, C.P. 72570 Puebla, Pue., M\'exico.}
\date{\today}
\begin{abstract}
We study flavor violating processes in the production or decay of
a neutral pseudoscalar meson $P^0$ in the framework of a type III
two Higgs Doublet Model with four-texture Yukawa couplings. We use
a version of the model where Yukawa interactions of neutral Higgs
bosons allow for flavor change at the tree-level, but conserves CP
symmetry. The physical Yukawa couplings respect CP-invariance due
to the Hermiticity conditions that we impose on the fermion mass
matrices. We focus on all possible $\tau^{\pm} \to l^{\pm}P^0 $
and $P^0 \to l^+l'^-$ decay channels, where $l,l'$ are charged
leptons. We find that these processes provide complementary
information on quark and lepton FCNC Yukawa couplings. In
particular flavor violating parameters in the quark sector,
$\chi_{sb}$ and $\chi_{db}$, are significantly constrained by
present experimental data, whereas the corresponding parameters in
the leptonic sector are less constrained.
\end{abstract}

\maketitle
\section{Introduction}

Flavor-changing neutral current (FCNC) processes and charge-parity
(CP) symmetry violation observed in the quark  sector are known to
be duly explained in the standard model (SM) framework via the quark
mixing mechanism \cite{ckm}. Nowadays, experimental searches of CP
violation and FCNC's in the leptonic sector are among the most
interesting problems in particle physics since they would eventually
shed some light on the origin of mixing and masses of leptons.
Actually, since LF violating  processes originating from mixing of
neutrinos are expected to occur at an unobservable small level
\cite{lfvmixneutrinos,lfv}, their observation would clearly indicate
evidence for New Physics beyond the SM \cite{standmod}.

In this paper, we study the LF violating processes that may occur in
the production or decay of a neutral pseudoscalar meson $P^0$ due to
the exchange of a neutral pseudoscalar Higgs boson ($A^0$). To be
more specific, we will focus on LF violation in all possible
$\tau^{\pm} \to l^{\pm}P^0 $ and $P^0 \to l^+l'^-$ decay channels
($l,l'$ are charged leptons) that are allowed by kinematics.
Although our main focus is on LF violating processes, for the
purposes of comparison we will also consider LF conserving decays of
neutral heavy mesons by assuming that they are induced by the Higgs
boson $A^0$ alone.

With the advent of B meson factories, large sets of $\tau$ lepton
pairs have been accumulated by both BaBar and Belle experiments
\cite{babarbelle} allowing to improve previous bounds on flavor
violating $\tau$ and $B$ meson decays. We note also that the
flavor violating decays of our interest have been studied in a
large variety of New Physics models and using different
approximations \cite{modelsa,modelsb}. A comprehensive set of
bounds on R-parity violating couplings from these LFV decays were
obtained in \cite{Dreiner:2006gu}. In this paper we study these
decays in the framework of a general two Higgs doublet model
type-III (2HDM-III) \cite{Barger:1989fj} which contains flavor
violation via Yukawa interactions of the neutral Higgs bosons at
the tree-level. In this paper we work in a type III 2HDM with
four-texture Yukawa couplings, where the Yukawa matrices are
assumed to be Hermitean, this implies that CP symmetry is
respected by the Yukawa interactions of neutral Higgs bosons which
largely simplifies the analysis since only the pseudoscalar Higgs
boson contributes to the decay amplitudes of the processes under
consideration. It is worth noticing that a restricted set of these
$\tau$ decays were considered in Ref.\cite{Li:2005rr} in the
2HDM-III but without assuming CP conservation.

This paper is organized as follows. In section 2, we provide the
Yukawa interactions for the neutral Higgs bosons. In section 3, we
derive the LFV effective four-fermion hamiltonian and we provide the
numerical values of couplings relevant for our calculations. The
formulae needed to evaluate the two-body decay rates are given in
section 4; from a comparison with the experimental upper limits on
branching fractions we derive the bounds on flavor
violating couplings. Finally, in section 5, we present our
conclusions.\\

\section{Flavor Violation in the 2HDM-III}

All lepton flavor violating processes in the SM have vanishing decay
probabilities at the tree level and even at 1-loop level they are
completely negligible \cite{lfvmixneutrinos} and beyond the reach of
present and planned experiments. Therefore, flavor violation at a
level accessible to experiments can be expected only in the
framework of New Physics scenarios. In this section we discuss the
Yukawa couplings of fermions in the 2HDM-III which can induce
flavor-changing neutral current processes both in the leptonic and
quark sectors. We also provide numerical values of couplings that
enter the hadronic matrix elements.

In the next subsections we shall derive the Yukawa interactions of
the scalar $(h^0, \, H^0)$ and pseudoscalar ($A^0$) neutral Higgs
bosons with the charged lepton and quark sectors of the 2HDM-III by
assuming four-texture mass matrices
\cite{DiazCruz:2004tr,DiazCruz:2004pj,GomezBock:2005hc}.

\subsection{The charged lepton sector}

The Yukawa lagrangian of the 2HDM-III for the charged lepton sector
is given by:
\begin{equation}
{\cal{L}}_Y^l = Y^{l}_{1ij}\bar{L_{i}}\Phi_{1}l_{Rj} +
Y^{l}_{2ij}\bar{L_{i}}\Phi_{2}l_{Rj} + h.c., \label{lagleptons}.
\end{equation}
where $\Phi_{1,2}=(\phi^+_{1,2}, \phi^0_{1,2})^T$ denote the Higgs
doublets and $L_i$ denote the doublet of left-handed leptons. The
specific choices for the Yukawa matrices $Y^l_{1,2}$ define the
versions of the 2HDM known as type I, II or III.

After spontaneous symmetry breaking the charged lepton mass matrix
is given by,
\begin{equation}
M_l= \frac{1}{\sqrt{2}}(v_{1}Y_{1}^{l}+v_{2}Y_{2}^{l}),
\end{equation}
Now we assume that both Yukawa matrices $Y^l_1$ and $Y^l_2$ have the
four-texture form and are Hermitian; following the conventions of
\cite{fourtext}, the lepton mass matrix is then written as: \beq
M_l= \left( \begin{array}{ccc}
0 & C_{l} & 0 \\
C_{l}^{*} & \tilde{B}_{l} & B_{l} \\
0 & B_{l}^{*} & A_{l}
\end{array}\right) ,
\eeq such that when $\tilde{B}_{l}\to 0$ one recovers the
six-texture form. We will also consider the following hierarchy,
$\mid A_{l}\mid \, \gg \, \mid \tilde{B}_{l}\mid,\mid B_{l}\mid
,\mid C_{l}\mid$, which is supported by the observed fermion masses
in the SM.

Because of the Hermiticity condition, both $\tilde{B}_{l}$ and
$A_{l}$ are real parameters, while the phases $\Phi_{B,C}$ of $C_l$
and $B_l$, can be removed from the mass matrix $M_l$ by defining:
$M_l=P^\dagger \tilde{M}_l P$, where $P={\rm diag}[1, e^{i\Phi_C},
e^{i(\Phi_B+\Phi_C)}]$, and the mass matrix $\tilde{M}_l$ includes
only the real parts of $M_l$. The  diagonalization of $\tilde{M}$ is
then obtained by means of an orthogonal matrix $O$, such that the
diagonal mass matrix is $\bar{M}_{l} = O^{T}\tilde{M}_{l}O$.

\bigskip

The lagrangian (\ref{lagleptons}) can be expanded in terms of the
mass-eigenstates for the neutral ($h^0,H^0,A^0$) and charged Higgs
bosons ($H^\pm$). The interactions of the neutral Higgs bosons are
then given by,
\begin{eqnarray}
{\cal{L}}_Y^{l} & = & \frac{g}{2}\left(\frac{m_{i}}{m_W}\right)
\bar{l}_{i}\left\{ \left[\frac{ \, \cos\alpha}{\cos\beta}\delta_{ij}+
\frac{\sqrt{2} \, \sin(\alpha - \beta)}{g \, \cos\beta}
\left(\frac{m_W}{m_{i}}\right)\tilde{Y}_{2ij}^{l}\right]H^{0} \right.
\nonumber \\
                 &  & \left. \ \ \ \ \ \ \ \ \ \ +
\left[-\frac{\sin\alpha}{\cos\beta} \delta_{ij}+ \frac{\sqrt{2} \,
\cos(\alpha - \beta)}{g \, \cos\beta}
\left(\frac{m_W}{m_{i}}\right)\tilde{Y}_{2ij}^{l}\right] h^{0} \right.
\nonumber \\
                 & & \left. \ \ \ \ \ \ \ \ \ \ + \, i
\left[-\tan\beta \delta_{ij}+  \frac{\sqrt{2} }{g \, \cos\beta}
\left(\frac{m_W}{m_{i}}\right)\tilde{Y}_{2ij}^{l}\right]
\gamma^{5}A^0 \right\} l_{j}  + h.c. \label{lagleptonsy}
\end{eqnarray}
where $\alpha$ denotes the mixing angle which is used to define the
physical mass eigenstates $h^0$ and $H^0$ in the CP-even Higgs
sector,and $\tan \beta$ is the ratio of vacuum expectation values of
the two Higgs doublets \cite{hixphen}. In the above expression, the
terms proportional to $\delta{ij}$ corresponds to the modification
of the 2HDM-II over the SM result, while the terms proportional to
$\tilde{Y}_2^l$ denotes the new contribution from the 2HDM-III.
Thus, the physical fermion-Higgs couplings respect CP-invariance,
despite the fact that the Yukawa matrices include complex phases;
this follows because of the Hermiticity conditions imposed on both
$Y_1^l$ and $Y_2^l$ matrices.

The correction terms to the lepton flavor conserving (LFC) and
violating (LFV) couplings, depend on the rotated matrix
$\tilde{Y}_{2}^{l} = O^{T}PY_{2}^{l}P^\dagger O$. We can derive
$\tilde{Y}_{2}^{l}$, by assuming that $Y_2^l$ has a four-texture
form, namely:
\beq Y_{2}^{l}= \left( \begin{array}{ccc}
0 & C_{2} & 0 \\
C_{2}^{*} & \tilde{B}_{2} & B_{2} \\
0 & B_{2}^{*} & A_{2}
\end{array}\right), \qquad
\mid A_{2}\mid \, \gg \, \mid \tilde{B}_{2}\mid,\mid B_{2}\mid ,\mid
C_{2}\mid. \eeq Since the orthogonal matrix that diagonalizes the
real matrix $\tilde{M}_{l}$ with the four-texture form, has the
form:
\beq O = \left( \begin{array}{ccc}
\sqrt{\frac{\lambda_{2}\lambda_{3}(A-\lambda_{1})}{A(\lambda_{2}-\lambda_{1})
(\lambda_{3}-\lambda_{1})}}& \eta \sqrt{\frac{\lambda_{1}\lambda_{3}
(\lambda_{2}-A)}{A(\lambda_{2}-\lambda_{1})(\lambda_{3}-\lambda_{2})}}
& \sqrt{\frac{\lambda_{1}\lambda_{2}(A-\lambda_{3})}{A(\lambda_{3}-
\lambda_{1})(\lambda_{3}-\lambda_{2})}} \\
-\eta
\sqrt{\frac{\lambda_{1}(\lambda_{1}-A)}{(\lambda_{2}-\lambda_{1})
(\lambda_{3}-\lambda_{1})}} &
\sqrt{\frac{\lambda_{2}(A-\lambda_{2})}
{(\lambda_{2}-\lambda_{1})(\lambda_{3}-\lambda_{2})}} & \sqrt{
\frac{\lambda_{3}(\lambda_{3}-A)}{(\lambda_{3}-\lambda_{1})(\lambda_{3}-
\lambda_{2})}} \\
\eta
\sqrt{\frac{\lambda_{1}(A-\lambda_{2})(A-\lambda_{3})}{A(\lambda_{2}
-\lambda_{1})(\lambda_{3}-\lambda_{1})}} &
-\sqrt{\frac{\lambda_{2}(A
-\lambda_{1})(\lambda_{3}-A)}{A(\lambda_{2}-\lambda_{1})(\lambda_{3}
-\lambda_{2})}} &
\sqrt{\frac{\lambda_{3}(A-\lambda_{1})(A-\lambda_{2})}
{A(\lambda_{3}-\lambda_{1})(\lambda_{3}-\lambda_{2})}}
\end{array}\right),
\eeq
where $m_{e}= m_{1} = \mid \lambda _{1}\mid, m_{\mu}= m_{2} =
\mid \lambda _{2}\mid, m_{\tau}= m_{3} = \mid \lambda _{3}\mid$ and
$\eta = \lambda_{2}/ m_{2}$, then the rotated form of $\tilde
{Y}_{2}^{l} $ acquires the general form,
\begin{eqnarray}
\tilde {Y}_{2}^{l}  & = & O^{T}PY_{2}^{l}P^{\dagger}O \nonumber \\
& = &\left( \begin{array}{ccc}
(\tilde {Y}_2^l)_{11}   & (\tilde {Y}_2^l)_{12}   & (\tilde {Y}_2^l)_{13}   \\
(\tilde {Y}_2^l)_{21}   & (\tilde {Y}_2^l)_{22}   & (\tilde {Y}_2^l)_{23}  \\
(\tilde {Y}_2^l)_{31}   & (\tilde {Y}_2^l)_{32}   & (\tilde
{Y}_2^l)_{33}
\end{array}\right).
\end{eqnarray}

The full expressions for the resulting entries of this matrix have a
complicated form. To derive a convenient approximation, we will
consider that the elements of the Yukawa matrix $Y_2^l$ exhibit the
same hierarchy as the full mass matrix, namely:
\begin{eqnarray}
C_{2} & = &  c_{2}\sqrt{\frac{m_{1}m_{2}m_{3}}{A}} \ ,  \\
B_{2} & = &  b_{2}\sqrt{(A - \lambda_{2})(m_{3}-A)} \ , \\
\tilde{B}_{2} & = & \tilde{b}_{2}(m_{3}-A + \lambda_{2}) \,  \\
A_{2} & = & a_{2}A.
\end{eqnarray}

In order to keep the same hierarchy for the elements of the mass
matrix, it is found that $A$ must satisfy the condition $ (m_{3}-
m_{2}) \leq A \leq m_{3}$. Thus, we propose the following relation
for $A$:
\begin{equation}
A  = m_{3}(1 -\beta z),
\end{equation}
where $z = m_{2}/m_{3} \ll 1$  and $0 \leq \beta \leq 1$.

If we now introduce the dimensionless matrix $\tilde{\chi}$, we get:
\begin{eqnarray}
\left( \tilde {Y}_{2}^{l} \right)_{ij}
&=& \frac{\sqrt{m_i m_j}}{v} \, \tilde{\chi}_{ij} \nonumber\\
&=&\frac{\sqrt{m_i m_j}}{v}\, {\chi}_{ij} \, e^{\vartheta_{ij}}\ .
\label{myleptons}
\end{eqnarray}
This expression differs from the usual Cheng-Sher $Ansatz$
\cite{chengsher} not only because of the appearance of the complex
phases, but also in the form of the real parts ${\chi}_{ij} =
|\tilde{\chi}_{ij}|$.

If we expand in powers of $z$, the entries of the
matrix $\tilde{\chi}$ become:
\begin{eqnarray}
\tilde{\chi}_{11} & = & [\tilde{b}_{2}-(c^*_{2}e^{i\Phi_{C}}
+c_{2}e^{-i\Phi_{C}} )]\eta
    +[a_{2}+\tilde{b}_{2}-(b^*_{2}e^{i\Phi_{B}} + b_{2}e^{-i\Phi_{B}} )]
         \beta \nonumber \\
\tilde{\chi}_{12} & = & (c_{2}e^{-i\Phi_{C}}-\tilde{b}_{2})
-\eta[a_{2}+ \tilde{b}_{2}-(b^*_{2}e^{i\Phi_{B}} +
b_{2}e^{-i\Phi_{B}} )] \beta ]
\nonumber \\
\tilde{\chi}_{13} & = & (a_{2}-b_{2}e^{-i\Phi_{B}}) \eta
\sqrt{\beta}
                           \nonumber  \\
\tilde{\chi}_{22}  & = & \tilde{b}_{2}\eta
+[a_{2}+\tilde{b}_{2}-(b^*_{2}e^{i\Phi_{B}} +b_{2}e^{-i\Phi_{B}} )]
         \beta \nonumber \\
\tilde{\chi}_{23} & = & (b_{2}e^{-i\Phi_{B}}-a_{2})
                              \sqrt{\beta} \nonumber  \\
\tilde{\chi}_{33} & = & a_{2} \label{chileptons}
\end{eqnarray}

As it can be easily checked, the diagonal elements
$\tilde{\chi}_{ii}$ are real and the phases appear in the
off-diagonal elements; we note that these elements are not
constrained by current low-energy phenomena. Furthermore the LFV
couplings satisfy some relations, such as: $|\tilde{\chi}_{23}| =
|\tilde{\chi}_{13}|$, which simplifies somehow the number of free
parameters.

Finally, in order to perform phenomenological studies it is found
convenient to rewrite the lagrangian given in Eq.
(\ref{lagleptonsy}) in terms of the $\tilde{\chi}_{ij}$'s as
follows:

\begin{eqnarray}
{\cal{L}}_Y^{l} & = & \frac{g}{2} \, \bar{l}_{i} \left[\left( \,
\frac{m_{i}}{m_W}\right)\frac{\cos\alpha}{\cos\beta} \, \delta_{ij}
+ \frac{\sin(\alpha - \beta)}{\sqrt{2} \, \cos\beta}
\left(\frac{\sqrt{m_i
m_j}}{m_W}\right)\tilde{\chi}_{ij}\right]l_{j}H^{0}
\\
                &   & + \frac{g}{2} \, \bar{l}_{i}
\left[-\left(\frac{m_{i}}{m_W}\right)\frac{\sin\alpha}{\cos\beta} \,
\delta_{ij} + \frac{\cos(\alpha - \beta)}{\sqrt{2} \, \cos\beta}
\left(\frac{\sqrt{m_i m_j}}{m_W}\right)\tilde{\chi}_{ij}\right]l_{j}
h^{0}
\nonumber \\
                &   & + \frac{ig}{2} \, \bar{l}_i
\left[-\left(\frac{m_i}{m_W}\right)\tan\beta \, \delta_{ij} +
\frac{1}{\sqrt{2} \, \cos\beta} \left(\frac{\sqrt{m_i
m_j}}{m_W}\right)\tilde{\chi}_{ij}\right] \gamma^{5}l_{j} A^0 +
h.c., \nonumber \label{leptons1}
\end{eqnarray}
where, unlike the Cheng-Sher {\it Ansatz}, $\tilde{\chi}_{ij}$ $(i
\neq j)$ are now complex parameters.

\subsection{The quark sector}

The Yukawa lagrangian for the quark fields in  the 2HDM-III model is
written as:
\begin{equation}
{\cal{L}}_Y^q = Y^{u}_1\bar{Q}_L \Phi_{1} u_{R} +
                   Y^{u}_2 \bar{Q}_L\Phi_{2}u_{R} +
Y^{d}_1\bar{Q}_L \Phi_{1} d_{R} + Y^{d}_2 \bar{Q}_L\Phi_{2}d_{R} +
h.c., \label{lagquarks}
\end{equation}
where $Q_L$ denote the doublet of left-handed quarks. Again, the
specific choices for the Yukawa matrices $Y^q_{1,2}$ ($q=u,d$)
define the versions of the 2HDM known as type I, II or III.

Following a similar procedure as in the case of the charged leptons
we can derive the Yukawa couplings for the physical neutral Higgs
bosons ($h^0,H^0,A^0$). If we introduce the dimensionless matrix
$\tilde{\chi}_{qq'}$:
\begin{eqnarray}
\left( \tilde {Y}_2^q \right)_{qq'}
&=& \frac{\sqrt{m_q m_{q'}}}{v} \, \tilde{\chi}_{qq'} \nonumber\\
&=&\frac{\sqrt{m_q m_{q'}}}{v}\, {\chi}_{qq'} \, e^{i
\vartheta_{qq'}} \label{myquarks}
\end{eqnarray}
to replace the Yukawa matrices in the rotated basis, we get the
explicit form of the Yukawa couplings for the neutral Higgs bosons:
\begin{eqnarray}
{\cal{L}}_Y^{q} & = & \frac{g}{2} \, \bar{d} \left[\left( \,
\frac{m_d}{m_W}\right)\frac{\cos\alpha}{\cos\beta} \, \delta_{dd\,'}
+ \frac{\sin(\alpha - \beta)}{\sqrt{2} \, \cos\beta}
\left(\frac{\sqrt{m_d
m_{d\,'}}}{m_W}\right)\tilde{\chi}_{dd\,'}\right]d\,'H^{0}
\\
                &   & + \frac{g}{2} \, \bar{d}
\left[-\left(\frac{m_d}{m_W}\right)\frac{\sin\alpha}{\cos\beta} \,
\delta_{dd\,'} + \frac{\cos(\alpha - \beta)}{\sqrt{2} \, \cos\beta}
\left(\frac{\sqrt{m_d
m_{d\,'}}}{m_W}\right)\tilde{\chi}_{dd\,'}\right]d\,'h^{0}
\nonumber \\
                &   & + \frac{ig}{2} \, \bar{d}
\left[-\left(\frac{m_d}{m_W}\right)\tan\beta \, \delta_{dd\,'} +
\frac{1}{\sqrt{2} \, \cos\beta} \left(\frac{\sqrt{m_d
m_{d\,'}}}{m_W}\right)\tilde{\chi}_{dd\,'}\right]
\gamma^{5} d\,' A^0 \nonumber \\
                &   & + \frac{g}{2} \, \bar{u}
\left[\left( \, \frac{m_u}{m_W}\right)\frac{\sin\alpha}{\sin\beta}
\, \delta_{uu'} - \frac{\sin(\alpha - \beta)}{\sqrt{2} \, \sin\beta}
\left(\frac{\sqrt{m_u
m_{u'}}}{m_W}\right)\tilde{\chi}_{uu'}\right]u'H^{0}
\nonumber \\
                &   & + \frac{g}{2} \, \bar{u}
\left[\left(\frac{m_u}{m_W}\right)\frac{\cos\alpha}{\sin\beta} \,
\delta_{uu'} - \frac{\cos(\alpha - \beta)}{\sqrt{2} \, \sin\beta}
\left(\frac{\sqrt{m_u
m_{u'}}}{m_W}\right)\tilde{\chi}_{uu'}\right]u'h^{0}
\nonumber \\
                &   & + \frac{ig}{2} \, \bar{u}
\left[-\left(\frac{m_u}{m_W}\right)\cot\beta \, \delta_{uu'} +
\frac{1}{\sqrt{2} \, \sin\beta} \left(\frac{\sqrt{m_u
m_{u'}}}{m_W}\right)\tilde{\chi}_{uu'}\right] \gamma^{5} u' A^0 +
h.c., \nonumber \label{quarks1}
\end{eqnarray}
where $d$ (respectively, $u$) stands for down-type (up-type) quarks.

In our discussion above, the flavor changing couplings
are contained in the entries $\tilde{\chi}_{ff'}$; these
dimensionless parameters are expected to be of order one or below
and we will consider that the bounds obtained from data are {\it
significant} if this turns out to be the case. In our calculations
below, we will neglect the contributions of $\tilde{\chi}_{ff}$ in
the flavor conserving vertices. Note that processes with flavor
violation at both vertices, like $\tau \to \mu K_S$ or $B^0 \to
e\mu$, are allowed to occur in our model. However, we will not
consider this type of decays because data allows to bound the
product of flavor violating couplings and not these couplings
themselves.

It is interesting to note that in the above version of the
2HDM-III, the assumption that the Yukawa matrices are Hermitean
leads to CP conservation, and this implies that the scalar
(pseudoscalar) Higgs bosons couple only to scalar (pseudoscalar)
fermionic currents. This will greatly simplify the analysis of
this paper since only the pseudoscalar Higgs boson will contribute
to the decays of our interest.

\section{Effective hamiltonian and coupling constants}

The Feynman graphs corresponding to the two-body decays of our
interest are mediated by a  pseudoscalar Higgs boson $A^0$ and are shown
in Figure 1.

\begin{figure}[hbt]
\begin{center}
\mbox{\includegraphics[width=12cm, height=6cm]{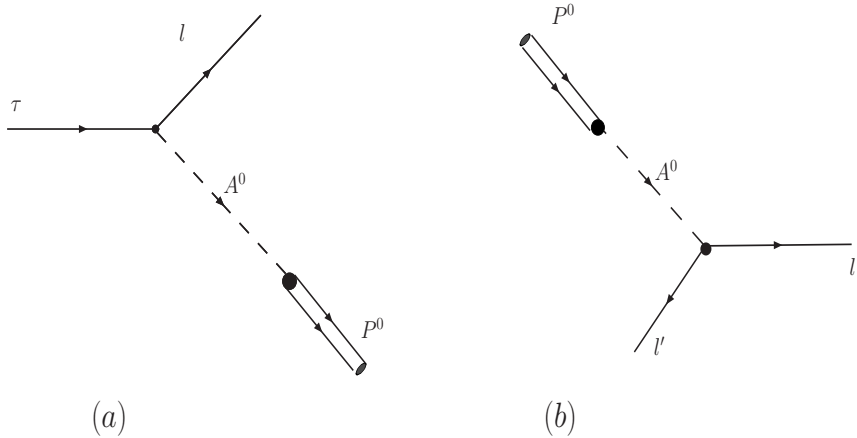}}
\caption[]{\label{fg:taufeynps} Two-body decays mediated by the
pseudoscalar Higgs boson $A^0$: (a) tau lepton decays and, (b)
Pseudoscalar meson decays.}
\end{center}
\end{figure}

In the local four-fermion approximation, which is valid for the energy
scales involved in the decays under consideration, it is more convenient
to work with the effective hamiltonian given by
\begin{equation}
\mathcal{H}=\sqrt{2}G_{F}\left(
\frac{m_{W}^{2}}{m_{A^0}^{2}}\right)g_{All'}\bar{\psi}_{l'}\gamma_{5}\psi_{l}
\times \left[\sum_{q,q'} g_{Aqq'}\bar{q'}\gamma_{5}q \right] +
h.c.
\end{equation}
where the coupling constants $g_{Aff'}$ can be easily identified
from the interaction lagrangians given in eqs. (1) and (2). The
above interaction hamiltonian allows us to cover all the cases for
flavor violation at the lepton and quark vertices as studied this
paper.

The relevant hadronic matrix elements that will enter in our
calculations are the following:
\begin{equation}
 (m_{q'}+m_q)\left< 0|i\bar{q'}\gamma_5 q|P\right>=h^q_{P} \ ,
\end{equation}
where $q, q'$ denote the quarks ($q'\not =q$ only when flavor violation
occurs at the meson vertex). Note that the $h^q_P$ constants
can be directly related to the usual pseudoscalar meson decay constants
$f_P$ (namely, $h_P=f_Pm_P^2$ for all the cases except the $\pi^0,
\eta,\eta'$ mesons).

In this work we will use the isospin limit where $m_u=m_d=3.3$ MeV.
For the other quark masses we will use \cite{Amsler:2008zzb}:
$m_s=0.104$, $m_c=1.27$, $m_b=4.2$ GeV. In our numerical
evaluations, we will use the following values of the decay constants
of the unflavored mesons \cite{Feldmann:1999uf} (here $q=u,d$):
\begin{eqnarray}
h^{q}_{\pi} &=&  2.38 \times 10^{-3}\ {\rm GeV}^3, \nonumber \\
h^{q}_{\eta}&=&2.0\times 10^{-3}\ {\rm GeV}^3,\nonumber \\
h^{q}_{\eta^{\prime}}&=&1.6\times 10^{-3}\ {\rm GeV}^3,\nonumber \\
h^{s}_{\eta}&=&-53\times 10^{-3}\ {\rm GeV}^3, \\
h^{s}_{\eta^{\prime}}&=& 65\times 10^{-3}\ {\rm GeV}^3.\nonumber
\end{eqnarray}
Equivalently, we can use information on the pseudoscalar decay
constants from Refs.
\cite{Amsler:2008zzb,Yao:2006px,Gray:2005ad,Wingate:2003gm}:
\begin{eqnarray}
f_{\pi^0}& =& (130\pm 5)\ \mbox{\rm MeV}, \  \nn \\
f_{K^0}& =& (155.5\pm 0.9)\ \mbox{\rm MeV},\ \nn \\
f_{D^0}& =& (205.8\pm 8.9)\ \mbox{\rm MeV}, \  \nn \\
f_{B^0}& =& (216\pm 22)\ \mbox{\rm MeV}, \  \nn \\
f_{B_s}& =& (260\pm 29)\ \mbox{\rm MeV} \ .
\end{eqnarray}
The quark content of unflavored mesons are the following:
\begin{eqnarray}
\pi^{0}&=&\frac{1}{\sqrt{2}}(\bar{u}u-\bar{d}d)\nonumber\\
\eta&=&\eta_{8}\cos\theta_P-\eta_{1}\sin\theta_P \nonumber\\
\eta^{\prime}&=&\eta_{8}\sin\theta_P+\eta_{1}\cos\theta_P
\label{pmesons}
\end{eqnarray}
\noindent where $\eta_{8,1}$ denote the octet and singlet isoscalar
mesons
and $\theta_P \approx -20^{\circ}$ is the mixing angle of pseudoscalar
mesons.

\section {Flavor violation in production/decay of $P^0$ mesons}

In this section we study the coupling of a pseudoscalar
meson $P^0$ to a  leptonic neutral current. We consider that the flavor
change can occur at either the hadronic or leptonic vertex. When we
compare our results with the current experimental bounds on branching
fractions, we   derive the constraints on the relevant flavor changing
Yukawa couplings of the $A^0$ Higgs boson.

The decay widths in the case of $\tau$ lepton decays are given by:
\begin{eqnarray}
\Gamma(\tau \rightarrow lP)&=&\frac{G_F^2}{8\pi}
\left(\frac{m_{W}}{m_{A^0}}\right)^{4}
\left[(m_{\tau}-m_{l})^{2}-m_{P}^2 \right]
\frac{\lambda^{1/2}(m_{\tau}^2, m_l^2, m_P^2)}{m_{\tau^3}} \nonumber  \\
&&\times g^{2}_{A\tau l}\left|\left<
P|\sum_{q,q'} g_{Aqq'}\bar{q'}\gamma_{5}q  |0\right>\right|^{2} \
,
\end{eqnarray}
while the corresponding widths for meson decays are:
\begin{eqnarray}
\Gamma(P\to ll')&=& \frac{G_F^2}{4\pi} \left(\frac{m_W}{m_{A^0}}
\right)^4 [m_P^2-(m_{l}-m_{l'})^2]\cdot \frac{\lambda^{1/2}(m_P^2,
m_l^2, m_{l'}^2)}{m_P^3} \nn \\
&& \ \times g_{All'}^2 \left|\langle 0|\sum_{q,q'}
g_{Aqq'}\ole{q'}\gamma_5 q|P\rangle \right|^2 \ .
 \end{eqnarray}
In the above expressions we have defined
$\lambda(x,y,z)=x^2+y^2+z^2-2xy-2xz-2yz$.\\

The hadronic matrix elements required for our numerical evaluations
are the following:
\begin{eqnarray}
\langle 0|O|\pi\rangle &=&\frac{i}{\sqrt{2}} \,
\frac{f_{\pi}m_{\pi}^2}{(m_u+m_d)}(g_{Auu}-g_{Add}) \nn \\
\langle 0|O|\eta \rangle & =&-\frac{i}{\sqrt{2}}\left[
(g_{Auu}+g_{Add})\frac{h_{\eta}^q}{(m_u+m_d)}+\sqrt{2}g_{Ass}
\frac{h_{\eta}^s}{2m_s} \right] \nn \\
\langle 0|O|\eta' \rangle & =&-\frac{i}{\sqrt{2}}\left[
(g_{Auu}+g_{Add})\frac{h_{\eta'}^q}{(m_u+m_d)}+\sqrt{2}g_{Ass}
\frac{h_{\eta'}^s}{2m_s} \right] \nn \\
\langle 0|O|D^0 \rangle &
=&-ig_{Auc} \, \frac{f_Dm_D^2}{m_c+m_u}  \nn \\
\langle 0|O|B^0 \rangle &
=&-ig_{Adb} \, \frac{f_Bm_B^2}{m_b+m_d}  \nn \\
\langle 0|O|B_s \rangle &
=&-ig_{Asb} \, \frac{f_{B_s}m_{B_s}^2}{m_b+m_s} \ ,
\end{eqnarray}
where the operator $O$ is defined by $O\equiv \sum_i
g_{Aij}\ole{q}_j\gamma_5 q_i$ and the values of quark masses and
decay constants were given in the previous section.

\noindent $g_{Aij}$ is the coupling of the Higgs neutral boson to a
fermion pair $ij$ and in our case are given as follows:

\begin{equation}
g_{All'} = \frac{ig}{2} \left[-\left(\frac{m_l}{m_W}\right)\tan\beta
\, \delta_{ll'} + \frac{1}{\sqrt{2} \, \cos\beta}
\left(\frac{\sqrt{m_l m_{l'}}}{m_W}\right)\tilde{\chi}_{ll'}\right],
\label{gall}
\end{equation}

\begin{equation}
g_{Auu'} = \frac{ig}{2} \left[-\left(\frac{m_u}{m_W}\right)\cot\beta
\, \delta_{uu'} + \frac{1}{\sqrt{2} \, \sin\beta}
\left(\frac{\sqrt{m_u m_{u'}}}{m_W}\right)\tilde{\chi}_{uu'}\right],
\label{gauu}
\end{equation}

\noindent and

\begin{equation}
g_{Add\,'} = \frac{ig}{2}
\left[-\left(\frac{m_d}{m_W}\right)\tan\beta \, \delta_{dd\,'} +
\frac{1}{\sqrt{2} \, \cos\beta} \left(\frac{\sqrt{m_d
m_{d\,'}}}{m_W}\right)\tilde{\chi}_{dd\,'}\right], \label{gadd}
\end{equation}

\noindent where $l,l'=e,\mu,\tau$; $u,u'=u,c,t$; and
$d,d\,'=d,s,b.$\\

In order to derive the bounds on the New Physics couplings, we use
the upper limits on the branching ratios reported by the PDG
\cite{Amsler:2008zzb}. In Tables \ref{tb:taudecay}, \ref{tb:pdecay1}
and \ref{tb:pdecay2}, we show our results as upper bounds on the
product of leptonic and quark couplings normalized to the square of
the Higgs boson $A^0$ for $\tau$ lepton and pseudoscalar meson
decays. These expressions look rather long, but they can be useful
to easily implement future updates. Tables \ref{tb:taudecay} and
\ref{tb:pdecay1} contain the upper bounds on LFV couplings, while
Table \ref{tb:pdecay2} refers to the bounds on the flavor changing
quark couplings. A comparison of results from Tables I and II shows
that experimental data on $\tau$ lepton decays highly constrain the
$\tau l$ LF couplings. Conversely we observe that decays of light
mesons provide very poor constraints on $\mu e$ LFV Yukawa
couplings, despite the tightly upper limits on their branching
ratios .

\begin{table}[hbt]
\begin{center}
\begin{tabular}{|c|cc|}
\hline
Process & BR (PDG 2008)\cite{Amsler:2008zzb} & Upper limit \\
\hline $\tau^- \to\pi^0 e^- $ & $< 8.0 \times 10^{-8}$ &
$\left|\frac{g_{A\tau e}}{m_{A^0}^2}\frac{1}{\sqrt{2}} \,
\frac{f_{\pi}m_{\pi}^2}{(m_u+m_d)}(g_{Auu}-g_{Add})\right|< 2.14\times 10^{-8}$ \\
 $\tau^- \to \pi^0 \mu^-$ & $< 1.1 \times 10^{-7}$ &
$\left|\frac{g_{A\tau \mu}}{m_{A^0}^2}\frac{1}{\sqrt{2}} \,
\frac{f_{\pi}m_{\pi}^2}{(m_u+m_d)}(g_{Auu}-g_{Add})\right|< 2.67\times 10^{-8}$ \\
 $\tau^- \to \eta e^-$ & $< 9.2 \times 10^{-8}$ & $
\left|\frac{g_{A\tau e}}{m_{A^0}^2}\frac{1}{\sqrt{2}}\left[
(g_{Auu}+g_{Add})\frac{h_{\eta}^q}{(m_u+m_d)}+\sqrt{2}g_{Ass}
\frac{h_{\eta}^s}{2m_s} \right]\right|<4.25\times 10^{-8}$ \\
$\tau^- \to \eta \mu^-$ & $< 6.5 \times 10^{-8}$ & $
\left|\frac{g_{A\tau \mu}}{m_{A^0}^2}\frac{1}{\sqrt{2}}\left[
(g_{Auu}+g_{Add})\frac{h_{\eta}^q}{(m_u+m_d)}+\sqrt{2}g_{Ass}
\frac{h_{\eta}^s}{2m_s} \right]\right|<2.27\times 10^{-8}$ \\
 $\tau^- \to \eta^{\prime} e^-$ & $< 1.6 \times 10^{-7}$ & $
\left|\frac{g_{A\tau e}}{m_{A^0}^2}\frac{1}{\sqrt{2}}\left[
(g_{Auu}+g_{Add})\frac{h_{\eta'}^q}{(m_u+m_d)}+\sqrt{2}g_{Ass}
\frac{h_{\eta'}^s}{2m_s} \right]\right|<4.24\times
10^{-8}$\\
 $\tau^- \to \eta^{\prime} \mu^-$ & $< 1.3 \times 10^{-7}$ & $
\left|\frac{g_{A\tau \mu}}{m_{A^0}^2}\frac{1}{\sqrt{2}}\left[
(g_{Auu}+g_{Add})\frac{h_{\eta'}^q}{(m_u+m_d)}+\sqrt{2}g_{Ass}
\frac{h_{\eta'}^s}{2m_s} \right]\right|<4.19\times
10^{-8}$\\
\hline
\end{tabular}
\caption[]{\label{tb:taudecay} LFV tau decays to a light
pseudoscalar mesons.}
\end{center}
\end{table}

\begin{table}[hbt]
\begin{center}
\begin{tabular}{|c|cc|}
\hline
Process & BR (PDG 2008)\cite{Amsler:2008zzb} & Upper limit \\
\hline $\pi^0\to \mu^+e^-$ & $< 3.8 \times 10^{-10}$ &
$\left|\frac{g_{Ae\mu}}{m_{A^0}^2}\frac{1}{\sqrt{2}} \,
\frac{f_{\pi}m_{\pi}^2}{(m_u+m_d)}(g_{Auu}-g_{Add})\right|<4.55\times
10^{-2}$ \\
 $\pi^0\to \mu^-e^+$ & $< 3.4 \times 10^{-9}$ &
$\left|\frac{g_{Ae\mu}}{m_{A^0}^2}\frac{1}{\sqrt{2}} \,
\frac{f_{\pi}m_{\pi}^2}{(m_u+m_d)}(g_{Auu}-g_{Add})\right|<0.14$ \\
 $\eta\to \mu^{\pm}e^{\mp}$ & $< 6.0 \times 10^{-6}$ & $
\left|\frac{g_{Ae\mu}}{m_{A^0}^2}\frac{1}{\sqrt{2}}\left[
(g_{Auu}+g_{Add})\frac{h_{\eta}^q}{(m_u+m_d)}+\sqrt{2}g_{Ass}
\frac{h_{\eta}^s}{2m_s} \right]\right|<14.81$ \\
 $\eta'\to \mu^{\pm} e^{\mp}$ & $< 4.7 \times 10^{-4}$ & $
\left|\frac{g_{Ae\mu}}{m_{A^0}^2}\frac{1}{\sqrt{2}}\left[
(g_{Auu}+g_{Add})\frac{h_{\eta'}^q}{(m_u+m_d)}+\sqrt{2}g_{Ass}
\frac{h_{\eta'}^s}{2m_s} \right]\right|<1.21\times 10^{3}$
\\
\hline
\end{tabular}
\caption[]{\label{tb:pdecay1} LFV decays of light pseudoscalar
mesons.}
\end{center}
\end{table}

\begin{table}[hbt]
\begin{center}
\begin{tabular}{|c|c c|}
\hline
Process & BR(PDG 2008)\cite{Amsler:2008zzb} & Upper limit \\
\hline
$B^0\to e^+e^-$& $<1.3\times 10^{-7}$&$\big|\frac{g_{Aee}}{m_{A^0}^2}g_{Adb} \, \frac{f_Bm_B^2}{m_b+m_d}\big|<9.17\times 10^{-10}$\\
$B^0\to \mu^+\mu^-$&$<1.5\times 10^{-8}$& $\big|\frac{g_{A\mu\mu}}{m_{A^0}^2}g_{Adb} \, \frac{f_Bm_B^2}{m_b+m_d}\big|<3.12\times 10^{-10}$\\
$B^0\to \tau^+\tau^-$&$< 4.1\times 10^{-3}$&$\big|\frac{g_{A\tau\tau}}{m_{A^0}^2}g_{Adb} \, \frac{f_Bm_B^2}{m_b+m_d}\big|<2.20\times 10^{-7}$\\
$B^0_s\to e^+e^-$&$<5,4\times 10^{-5}$&$\big|\frac{g_{Aee}}{m_{A^0}^2}g_{Asb} \, \frac{f_{B_s}m_{B_s}^2}{m_b+m_s}\big|<1.86\times 10^{-8}$\\
$B^0_s\to \mu^+\mu^-$&$<4.7\times 10^{-8}$&$\big|\frac{g_{A\mu\mu}}{m_{A^0}^2}g_{Asb} \, \frac{f_{B_s}m_{B_s}^2}{m_b+m_s}\big|<5.49\times 10^{-10}$\\
$D^0\to e^+e^-$&$<1.2\times 10^{-6}$&$\big|\frac{g_{Aee}}{m_{A^0}^2}g_{Auc} \, \frac{f_Dm_D^2}{m_c+m_u}\big|<2.56\times 10^{-8}$\\
$D^0\to \mu^+\mu^-$&$<1.3\times 10^{-6}$&$\big|\frac{g_{A\mu\mu}}{m_{A^0}^2}g_{Auc} \, \frac{f_Dm_D^2}{m_c+m_u}\big|<2.68\times 10^{-8}$\\
\hline
\end{tabular}
\caption[]{\label{tb:pdecay2} Flavor violation in decays of
pseudoscalar mesons.}
\end{center}
\end{table}

\begin{figure}[hbt]
\begin{center}
\mbox{\includegraphics[width=12cm, height=8cm]{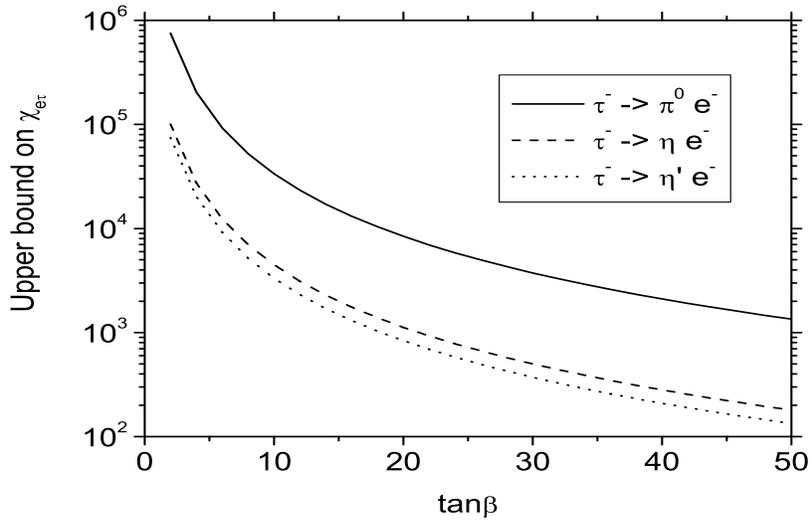}}
\caption[]{\label{fg:chi13} Upper bound on $\chi_{e\tau}$ as a
function of $\tan\beta$ for $m_{A^0}=300$ GeV (See Table
\ref{tb:taudecay}).}
\end{center}
\end{figure}

\begin{figure}[hbt]
\begin{center}
\mbox{\includegraphics[width=12cm, height=8cm]{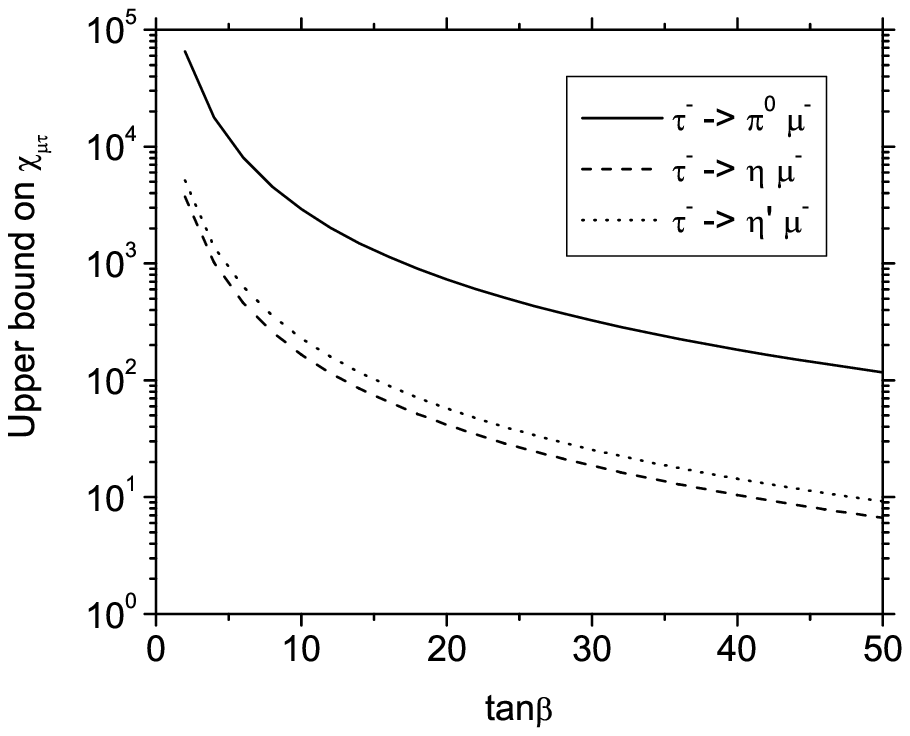}}
\caption[]{\label{fg:chi23} Upper bound on $\chi_{\mu\tau}$ as a
function of $\tan\beta$ for $m_{A^0}=300$ GeV (See Table
\ref{tb:taudecay}).}
\end{center}
\end{figure}

\begin{figure}[hbt]
\begin{center}
\mbox{\includegraphics[width=12cm, height=8cm]{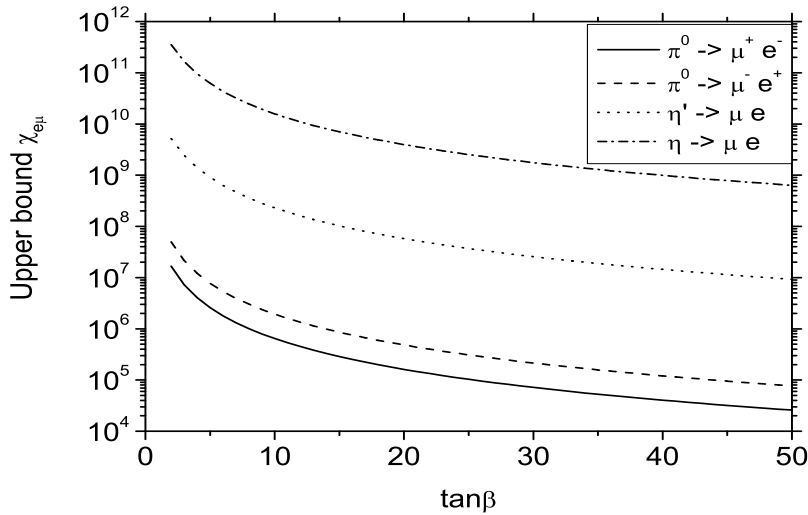}}
\caption[]{\label{fg:chi12} Upper bound on $\chi_{e\mu}$ as a
function of $\tan\beta$ for $m_{A^0}=300$ GeV (See Table
\ref{tb:pdecay1}).}
\end{center}
\end{figure}

Using the definitions of the effective couplings $g_{Aff'}$ and the
approximations noticed in section II we can translate the bounds
shown in Tables I--III on the dimensionless couplings
$\tilde{\chi}_{ff'}$. As we mentioned before, we will say that a
constraint is significant if these dimensionless couplings turn out
to be smaller that unity. In Figures \ref{fg:chi13}, \ref{fg:chi23},
\ref{fg:chi12} we show the bounds on the Yukawa couplings
$\chi_{e\tau}=|\tilde{\chi}_{e\tau}|$,
$\chi_{\mu\tau}=|\tilde{\chi}_{\mu\tau}|$,
$\chi_{e\mu}=|\tilde{\chi}_{e\mu}|$, as a function of $\tan\beta$ by
taking $m_{A^0}=300$ GeV, respectively. As we have mentioned in
section II, our imposition of the Hermiticity conditions on the
fermion mass matrices which implies CP-conservation in the Yukawa
sector allows us to derive bounds directly on the couplings of the
pseudoscalar Higgs boson. As we observe in Figs.
\ref{fg:chi13}--\ref{fg:chi12}, these bounds become more restrictive
when $\tan\beta$ increases. Conversely, these bounds are relaxed for
increasing values of the $A^0$ Higgs boson mass because these bounds
are proportional to $m_{A^0}^2$. It is shown in Fig. \ref{fg:chi13}
that the most restrictive bound on $\chi_{e\tau}$ is obtained from
$\tau^{-} \to \eta^{\prime} e^-$ decay, where we obtain
$\chi_{e\tau}<134 $ for $m_{A^0}=300$ GeV and $\tan\beta=50$. From
Fig. \ref{fg:chi23} we observe that the most restrictive bound on
$\chi_{\mu\tau}$ is obtained from $\tau^{-} \to \eta \mu^-$ decay,
with $\chi_{\mu\tau}<6.68$ for $m_{A^0}=300$ GeV and $\tan\beta=50$.
We recall here that $\chi_{e\tau}=\chi_{\mu\tau}$ in the model under
consideration \cite{DiazCruz:2004tr,DiazCruz:2004pj}.

\begin{figure}[hbt]
\begin{center}
\mbox{\includegraphics[width=12cm, height=8cm]{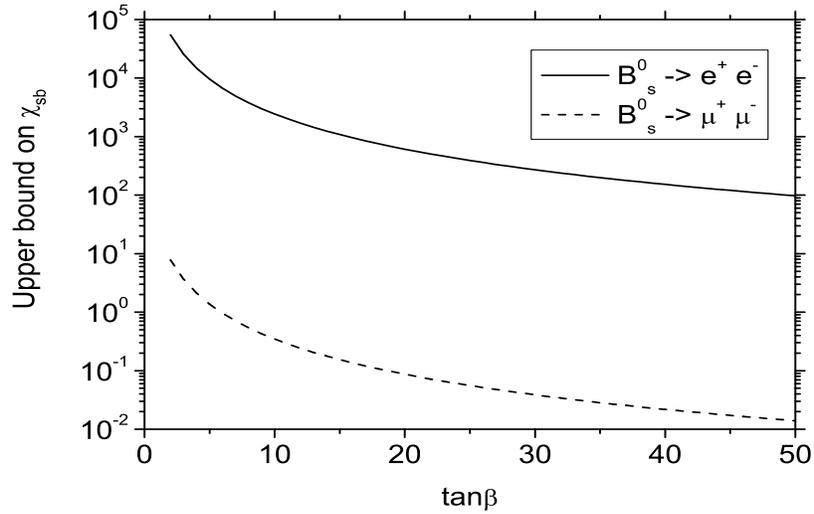}}
\caption[]{\label{fg:chisb} Upper bound on $\chi_{sb}$ as a function
of $\tan\beta$ for $m_{A^0}=300$ GeV (See Table \ref{tb:pdecay2}).}
\end{center}
\end{figure}

\begin{figure}[hbt]
\begin{center}
\mbox{\includegraphics[width=12cm, height=8cm]{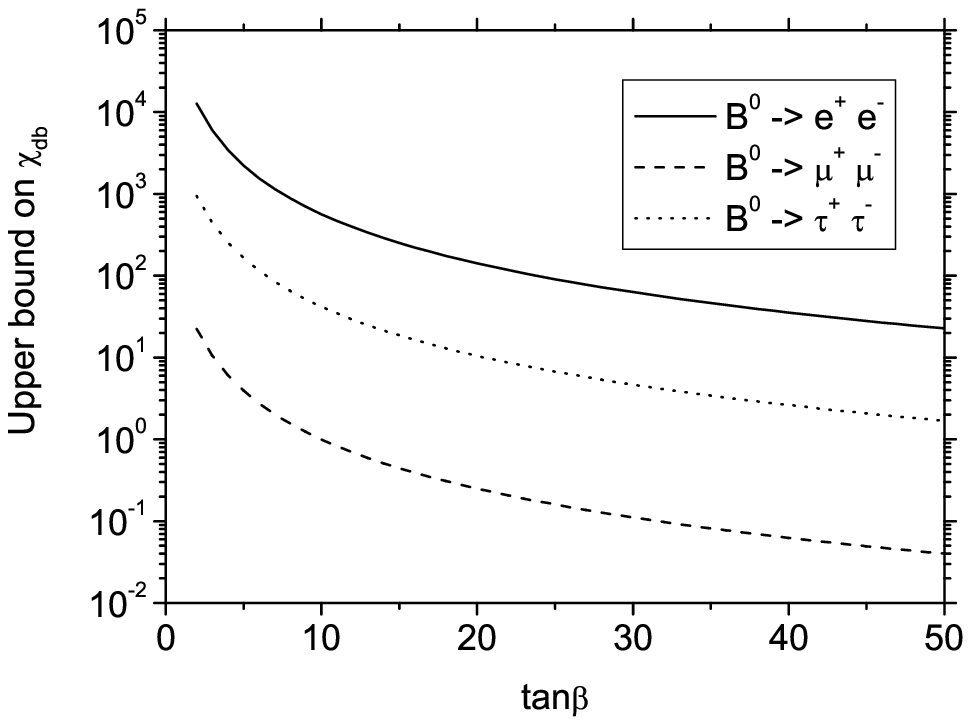}}
\caption[]{\label{fg:chidb} Upper bound on $\chi_{db}$ as a function
of $\tan\beta$ for $m_{A^0}=300$ GeV (See Table \ref{tb:pdecay2}).}
\end{center}
\end{figure}

\begin{figure}[hbt]
\begin{center}
\mbox{\includegraphics[width=12cm, height=8cm]{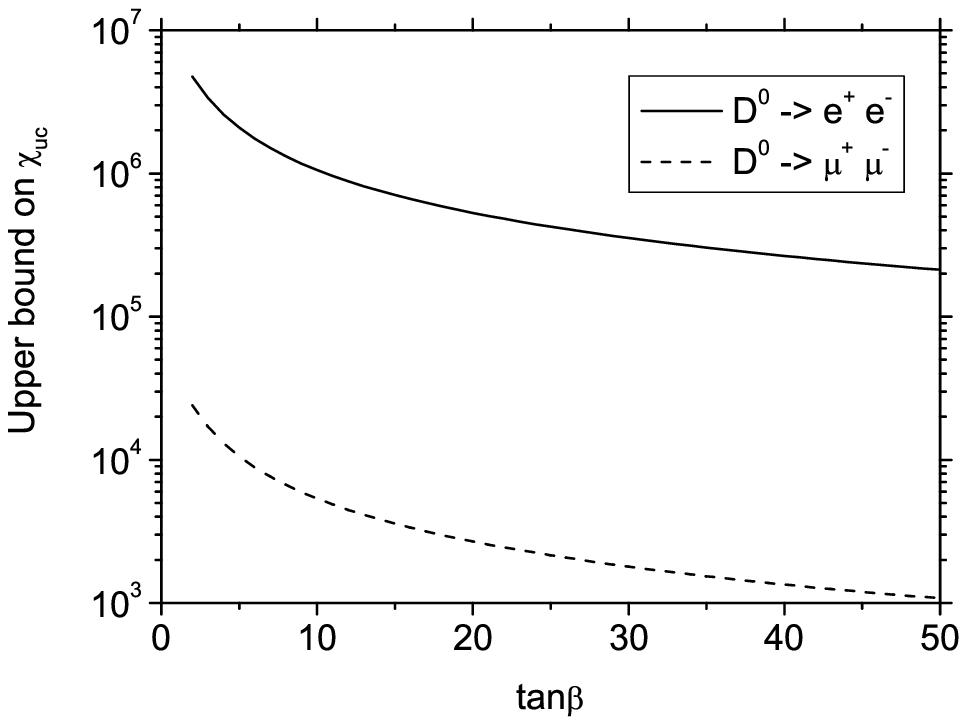}}
\caption[]{\label{fg:chiuc} Upper bound on $\chi_{uc}$ as a function
of $\tan\beta$ for $m_{A^0}=300$ GeV (See Table \ref{tb:pdecay2}).}
\end{center}
\end{figure}

In Figures \ref{fg:chisb}--\ref{fg:chiuc} we show the bounds on the
Yukawa couplings $\chi_{sb}=|\tilde{\chi}_{sb}|$,
$\chi_{db}=|\tilde{\chi}_{db}|$, $\chi_{uc}=|\tilde{\chi}_{uc}|$ as
a function of $\tan\beta$ by taking $m_{A^0}=300$ GeV, respectively.
As it can be seen in Fig. \ref{fg:chisb}, the most restrictive bound
on $\chi_{sb}$ is obtained from $B^0_{s} \to \mu^- \mu^+$ decay,
where we obtain $\chi_{sb}<1.40 \times 10^{-2}$ for $m_{A^0}=300$
GeV and $\tan\beta=50$. We observe in Fig. \ref{fg:chidb} that the
most restrictive bound on $\chi_{db}$ is obtained from $B^0 \to
\mu^- \mu^+$ decay, we obtain $\chi_{db}<4.01 \times 10^{-2}$ for
$m_{A^0}=300$ GeV and $\tan\beta=50$. We recall here that
$\chi_{db}=\chi_{sb}$ in the model under consideration. Using the
same input parameters for $\tan \beta$ and $m_{A^0}$, we observe in
Fig. \ref{fg:chiuc} that the most restrictive bound on $\chi_{uc}$
is obtained from $D^0 \to \mu^- \mu^+$ decay, namely
$\chi_{uc}<O(10^3) $ which is indeed very poor. We should note that
the upper bounds on the $\chi_{ff'}$ quark couplings are rather
conservative as long as the SM will
also give a contribution via the usual quark mixing  mechanism.\\

In the context of the same model used in this article, in Ref.
\cite{DiazCruz:2004pj} we have reported bounds on the $\chi_{sb}$
coupling, which are of the same order of magnitude than those
presented here. However, we want to point out that the upper bounds
reported in the present work have been obtained by using processes
which involve only $A^0$-exchange contributions, while those
reported in Ref. \cite{DiazCruz:2004pj} have been gotten by using
processes which involve $h^0$-exchange contributions. Hence, our
bounds depend only on $\tan\beta$, while the bounds reported in Ref.
\cite{DiazCruz:2004pj} depend on both parameters, $\tan\beta$ and
$\alpha$.

Finally, we can compare processes involving flavor violation at one
and the two vertices by taking the ratio of the $R_{ll'}\equiv
\Gamma(P^0 \to l^+l'^{-})/\Gamma(P^0 \to l^+l^-)$ decay rates (here
$l \not = l'$). This ratio is independent of the hadronic parameters
and of the $A^0$ boson mass, actually $R_{ll'} \approx
(m_{l'}/2m_l)\cdot (\chi_{ll'}/\sin \beta)^2$ for heavy meson decays
into light leptons. Thus, for large enough values of $\tan \beta$
(typically $\tan \beta \geq 5$) and under the assumption that
$\chi_{ll'} \leq 1$, $R_{\mu e}$ gets suppressed by a least a factor
of the $e/\mu$ mass ratio.

\section{Conclusions}

In this paper we have studied the lepton flavor violation induced
by the Yukawa couplings of neutral pseudoscalar Higgs boson $A^0$
of the 2HDM-III in the two-body decays of $\tau$ leptons and
pseudoscalar mesons. Under the assumption that the Yukawa matrices
are Hermitean which implies that CP is conserved by these
interactions, we are able to get bounds on the flavor-violating
Yukawa couplings of the pseudoscalar Higgs boson. Using present
data we have found the strongest bounds for the lepton-flavor
violating couplings in the case of $\tau$ lepton decays involving
($\eta,\ \eta'$) mesons. Clearly, improved experimental upper
limits on these decays by one or two orders of magnitude will
produce significant bounds on the LF violating couplings. We have
also considered LF conserving (but FCNC quark coupling) decays of
neutral mesons. In this case we are able to find very significant
constraints on the flavor-changing quark couplings under the
conservative assumption that the SM contribution is negligible. In
particular, we find $\chi_{sb}<1.40 \times 10^{-2}$ and
$\chi_{db}<4.01 \times 10^{-2}$ for the typical values
$m_{A^0}=300$ GeV and $\tan\beta=50$. We thus conclude that
neutral pseudoscalar mesons produced in $\tau$ lepton two-body
decays and their decays to charged lepton pairs, provide
information on the flavor-violating couplings that is
complementary to the ones from the CP-even Higgs bosons.

\begin{center}
{\bf ACKNOWLEDGMENTS}
\end{center}

This work was supported in part by {\it Consejo Nacional de Ciencia
y Tecnolog\'{\i}a} (M\'exico).

\begin{thebibliography}{99}

\bibitem{ckm}
  N.~Cabibbo,
  Phys.\ Rev.\ Lett.\  {\bf 10}, 531 (1963); M.~Kobayashi and
T.~Maskawa,  Prog.\ Theor.\ Phys.\  {\bf 49}, 652 (1973).
\bibitem{lfvmixneutrinos}
  S.~Bergmann and Y.~Grossman,
  Phys.\ Rev.\  D {\bf 59}, 093005 (1999)
  [arXiv:hep-ph/9809524];
  S.~Bergmann, Y.~Grossman and D.~M.~Pierce,
  Phys.\ Rev.\  D {\bf 61}, 053005 (2000)
  [arXiv:hep-ph/9909390];
  A.~Masiero, S.~K.~Vempati and O.~Vives,
  New J.\ Phys.\  {\bf 6}, 202 (2004)
  [arXiv:hep-ph/0407325];
  G.~Altarelli, F.~Feruglio and I.~Masina,
  Phys.\ Lett.\  B {\bf 472}, 382 (2000)
  [arXiv:hep-ph/9907532].

\bibitem{lfv}
  J.~A.~Casas and A.~Ibarra,
  arXiv:hep-ph/0109161.
  W.~Rodejohann,
  Phys.\ Rev.\  D {\bf 62} (2000) 013011
  [arXiv:hep-ph/0003149].
  V.~D.~Barger,
  arXiv:hep-ph/0102052.

\bibitem{standmod} S. Weinberg, Phys. Rev. Lett. {\bf 19} (1967) 1264; A. Salam,
in Proceedings of the 8th Nobel Symposium (Stockholm) , edited by N.
Svartholm, (Almqvist and Wiksell, Stockholm, 1968) p. 367; S.L.
Glashow, J. Illiopoulos and L. Maiani, Phys. Rev. {\bf D2} (1970)
1285

\bibitem{babarbelle}
URL addresses: http://www-public.slac.stanford.edu/babar,
http://belle.kek.jp/

\bibitem{modelsa}
  A.~Ilakovac and A.~Pilaftsis,
  Nucl.\ Phys.\  B {\bf 437}, 491 (1995)
  [arXiv:hep-ph/9403398];
  T.~Fukuyama, A.~Ilakovac and T.~Kikuchi,
  Eur.\ Phys.\ J.\  C {\bf 56}, 125 (2008)
  [arXiv:hep-ph/0506295];
  G.~Cvetic, C.~Dib, C.~S.~Kim and J.~D.~Kim,
  Phys.\ Rev.\  D {\bf 66}, 034008 (2002)
  [Erratum-ibid.\  D {\bf 68}, 059901 (2003)]
  [arXiv:hep-ph/0202212];
  X.~Y.~Pham,
  Eur.\ Phys.\ J.\  C {\bf 8}, 513 (1999)
  [arXiv:hep-ph/9810484];
  S.~Fajfer and A.~Ilakovac,
  Phys.\ Rev.\  D {\bf 57}, 4219 (1998);
  M.~J.~Herrero, J.~Portoles and A.~M.~Rodriguez-Sanchez,
  arXiv:0903.5151 [hep-ph];
  E.~Arganda, M.~J.~Herrero, J.~Portoles, A.~Rodriguez-Sanchez and A.~M.~Teixeira,
  AIP Conf.\ Proc.\  {\bf 1078}, 335 (2009)
  [arXiv:0810.0163 [hep-ph]];
  E.~Arganda, M.~J.~Herrero and J.~Portoles,
  JHEP {\bf 0806}, 079 (2008)
  [arXiv:0803.2039 [hep-ph]];
  E.~Arganda and M.~J.~Herrero,
  Phys.\ Rev.\  D {\bf 73}, 055003 (2006)
  [arXiv:hep-ph/0510405];
  S.~Antusch, E.~Arganda, M.~J.~Herrero and A.~M.~Teixeira,
  Nucl.\ Phys.\ Proc.\ Suppl.\  {\bf 169}, 155 (2007)
  [arXiv:hep-ph/0610439];
  Z.~H.~Li, Y.~Li and H.~X.~Xu,
  arXiv:0901.3266 [hep-ph];
  C.~X.~Yue, L.~H.~Wang and W.~Ma,
  Phys.\ Rev.\  D {\bf 74}, 115018 (2006)
  [arXiv:hep-ph/0611054];
  W.~j.~Li, Y.~d.~Yang and X.~d.~Zhang,
  Phys.\ Rev.\  D {\bf 73}, 073005 (2006)
  [arXiv:hep-ph/0511273].

\bibitem{modelsb}
  W.~Skiba and J.~Kalinowski,
  Nucl.\ Phys.\  B {\bf 404}, 3 (1993);
  J.~L.~Hewett, S.~Nandi and T.~G.~Rizzo,
  Phys.\ Rev.\  D {\bf 39}, 250 (1989);
  H.~E.~Logan and U.~Nierste,
  Nucl.\ Phys.\  B {\bf 586}, 39 (2000)
  [arXiv:hep-ph/0004139];
  G.~Lopez Castro, R.~Martinez and J.~H.~Munoz,
  Phys.\ Rev.\  D {\bf 58}, 033003 (1998)
  [arXiv:hep-ph/9804368];
  C.~S.~Huang, W.~Liao, Q.~S.~Yan and S.~H.~Zhu,
  Phys.\ Rev.\  D {\bf 63}, 114021 (2001)
  [Erratum-ibid.\  D {\bf 64}, 059902 (2001)]
  [arXiv:hep-ph/0006250];
  R.~A.~Diaz, R.~Martinez and C.~E.~Sandoval,
  Eur.\ Phys.\ J.\  C {\bf 41}, 305 (2005)
  [arXiv:hep-ph/0406265].
\bibitem{Dreiner:2006gu}
  H.~K.~Dreiner, M.~Kramer and B.~O'Leary,
  Phys.\ Rev.\  D {\bf 75}, 114016 (2007).

\bibitem{Barger:1989fj}
  V.~D.~Barger, J.~L.~Hewett and R.~J.~N.~Phillips,
  Phys.\ Rev.\  D {\bf 41}, 3421 (1990).

\bibitem{Li:2005rr}
  W.~j.~Li, Y.~d.~Yang and X.~d.~Zhang,
  Phys.\ Rev.\  D {\bf 73}, 073005 (2006)
  [arXiv:hep-ph/0511273].

\bibitem{DiazCruz:2004tr}
  J.~L.~Diaz-Cruz, R.~Noriega-Papaqui and A.~Rosado,
  Phys.\ Rev.\  D {\bf 69}, 095002 (2004).

\bibitem{DiazCruz:2004pj}
  J.~L.~Diaz-Cruz, R.~Noriega-Papaqui and A.~Rosado,
  Phys.\ Rev.\  D {\bf 71}, 015014 (2005).

\bibitem{GomezBock:2005hc}
  M.~Gomez-Bock and R.~Noriega-Papaqui,
  J.\ Phys.\ G {\bf 32}, 761 (2006).

\bibitem{fourtext} H. Fritzsch and Z. Z. Xing, Phys. Lett. B {\bf 555}, 63
(2003) (arXiv: hep-ph/0212195).

\bibitem{hixphen} For a review, see J.F. Gunion, H.E. Haber, G.L. Kane and S. Dawson,
{\it The Higgs Hunter's Guide} (Addison-Wesley Publishing Company,
1990).

\bibitem{chengsher} T.P. Cheng and M. Sher,
Phys.\ Rev.\ D {\bf 35} (1987) 3484.

\bibitem{Amsler:2008zzb}
  C.~Amsler {\it et al.}  [Particle Data Group],
  Phys.\ Lett.\  B {\bf 667}, 1 (2008).

\bibitem{Feldmann:1999uf}
  T.~Feldmann,
  Int.\ J.\ Mod.\ Phys.\  A {\bf 15}, 159 (2000).

\bibitem{Yao:2006px}
  W.~M.~Yao {\it et al.}  [Particle Data Group],
  J.\ Phys.\ G {\bf 33}, 1 (2006).

\bibitem{Gray:2005ad}
  A.~Gray {\it et al.}  [HPQCD Collaboration],
  Phys.\ Rev.\ Lett.\  {\bf 95}, 212001 (2005).

\bibitem{Wingate:2003gm}
  M.~Wingate, C.~T.~H.~Davies, A.~Gray, G.~P.~Lepage and J.~Shigemitsu,
  Phys.\ Rev.\ Lett.\  {\bf 92}, 162001 (2004).

\end{thebibliography}
\end{document}